\documentclass[reprint,aps,amsmath,nofootinbib]{revtex4-2}

\usepackage{dcolumn}
\usepackage{hyperref}
\usepackage{color}

\everymath{\displaystyle}

\begin{document}

\title{Analytic structure of stress-energy response functions and new Kubo formulae}

\author{Sangyong Jeon}
\email{sangyong.jeon@mcgill.ca}
\affiliation{Department of Physics, McGill University, 3600 rue University, Montreal,
H3A2T8 QC, Canada}
\author{Alina Czajka}
\email{alina.czajka@ncbj.gov.pl}
\affiliation{Theoretical Physics Division, National Centre for Nuclear Research, Pasteura 7, Warsaw 02-093, Poland}
\author{Juhee Hong}
\email{juhehong@gmail.com}
\affiliation{Department of Physics and Institute of Physics and Applied Physics, Yonsei University, Seoul 03722, Korea}

\date{\today}

\begin{abstract}

Determining the transport properties of Quark-Gluon Plasma is one of the most important
aspects of relativistic heavy ion collision studies.
Field-theoretical calculations of the transport coefficients such as the shear and bulk
viscosities require Kubo formulae which in turn require real-time
correlation functions of stress-energy tensors.
Consequently, knowing the analytic structure of these correlation functions is 
essential in any such studies.
Using the energy-conservation laws and the results from the gravity-hydrodynamics
analysis, we determine the low-frequency and low-wavenumber analytic structures of all
stress-energy correlation functions in the rest frame of the medium. 
By comparing with the diffusion and sound spectra from the second-order and the third-order
relativistic hydrodynamics,
various new Kubo formulae are derived in the limit where the zero-frequency limit is taken
first.
We also show that the meaning of the Kubo formulae for relaxation times can change when higher-order terms are
added to hydrodynamics. A subtle issue of taking the zero frequency and
zero wavenumber limits when using skeleton diagrams is addressed as well. 
\end{abstract}

\maketitle

\section{Introduction}

The main purpose of studying relativistic heavy ion collisions is to determine the
properties of Quark-Gluon Plasma (QGP).
Among the QGP properties, the equation of state and the transport coefficients are 
the most studied in relativistic heavy ion collisions. 
The QGP equation of state can be reliably calculated on the lattice up to the electroweak
scale as it only involves the calculation of the stress-energy tensor average
\cite{Borsanyi:2013bia,HotQCD:2014kol,Bazavov:2017dsy,Bresciani:2025vxw}. 
Field theoretical calculations of transport coefficients, however, are more complex
as they need to use the Kubo formulae involving real-time correlation functions.

Kubo formulae have been a staple of theoretical physics ever since it was first derived
\cite{doi:10.1143/JPSJ.12.1203,
doi:10.1143/JPSJ.12.570,
Kadanoff:1963axw}.
Relativistic field theory calculations of viscosities using the Kubo
formulae have a correspondingly long history
\cite{Hosoya:1983id,Hosoya:1983xm,
Karsch:1986cq,
Jeon:1994if,
Jeon:1995zm,Wang:1999gv,Carrington:1999bw,Wang:2002nba,
Arnold:2000dr,Arnold:2003zc,
Nakamura:2004sy,
Arnold:2006fz,
Huot:2006ys,Meyer:2007dy,Meyer:2007ic,Meyer:2011gj,
Gagnon:2007qt,Liu:2008az,Carrington:2009xf,Carrington:2009kh,
Lu:2011df,Chen:2012jc,Lang:2013lla,Christiansen:2014ypa,
Jeon:2015dfa,
Francis:2015daa,
Czajka:2017bod,Czajka:2017wdo,Harutyunyan:2017ttz,Harutyunyan:2017ieu,
Astrakhantsev:2017nrs,
Czajka:2018bod,Czajka:2018egm,Ghiglieri:2018dib,
Astrakhantsev:2018oue,
Romatschke:2021imm,
Altenkort:2022yhb,
Danhoni:2024ewq,
PavanPavan:2025hzr,
Kovtun:2004de,Buchel:2007mf}.
The standard Kubo formulae for the shear and the bulk viscosities can be derived 
by considering
the analytic structure of stress-energy tensor
correlation functions in the
the small-wavenumber ($k = |{\bf k}|$) and small-frequency ($\omega$) limit.
Such derivations so far (for instance, see \cite{Jeon:1994if,Jeon:2015dfa}) have used a
limited amount of information on how the correlation functions depend on the
transport coefficients. 

Recently, more information on the structure of the correlation functions 
became available due to the new developments in the gravity-hydrodynamic analysis 
\cite{Baier:2007ix,
York:2008rr,
Hong:2010at,
Moore:2010bu,
Moore:2012tc,
Romatschke:2015gic,
Kleinert:2016nav,
Romatschke:2017ejr,
Romatschke:2019gck,
Weiner:2022kgx}
as well as the third order relativistic hydrodynamics
\cite{deBrito:2023tgb,
Brito:2021iqr,
Jaiswal:2013vta,
Panday:2024hqp,
El:2009vj,
Ye:2024phs,
Younus:2019rrt,
Chattopadhyay:2014lya,
Diles:2019uft,
Diles:2023tau,
Grozdanov:2015kqa}.
To take advantage of the new information,
one needs to understand the analytic structure of
the stress-energy tensor correlation functions
more thoroughly.
The main purpose of this work is to determine 
the general analytic structure of the correlation functions
that are compatible with the new information from the gravity-hydrodynamics
and the third-order hydrodynamics.
By doing so, we are able to derive a number of new Kubo-formulae for the viscosities as we will show
below.
Interestingly, in most of the new formulae, the order of the limits is reversed. 
That is, the zero $\omega$ limit is taken first instead of the zero $k$ limit.

Schematically, the usual Kubo formula takes the following form
\begin{align}
\gamma_A & =
iC_A \lim_{\omega\to 0}\lim_{k\to 0}
\partial_\omega G_R^{\hat{A}\hat{A}}(\omega, {\bf k})
\label{eq:schematic_Kubo}
\end{align}
where $\hat{A}$ is the operator that is coupled to an external force that introduces
a perturbation to the underlying equilibrium system.
The retarded correlation function $G_R^{\hat{A}\hat{A}}(\omega, {\bf k})$ is calculated in
equilibrium, and $\gamma_A$ is the transport coefficient that controls the speed
of dissipation of the perturbation. The coefficient $C_A$ is an $O(1)$ constant.

The $\omega$-derivative is present
because the Kubo formula above is actually about the small $\omega$ behaviour
of the spectral density
defined as twice the imaginary part of the retarded correlation
function. 
The order of the limits in Eq.(\ref{eq:schematic_Kubo}) corresponds to the thermodynamic
limit where the zero $k$ limit (equivalently, the infinite volume limit)
is taken before the zero $\omega$ limit.
The two limits usually do not commute.

There was, however, an exception.
In
Ref.\cite{Baier:2007ix}
and many subsequent papers
\cite{Romatschke:2015gic,
Romatschke:2019gck,
Weiner:2022kgx,
Romatschke:2017ejr,
Moore:2012tc}
one finds the following
expansion of a retarded correlation function of the stress-energy tensor component 
$\hat{T}^{xy}$ 
\begin{align}
G_R^{xy,xy}(\omega, k_z)
= -i\eta\omega + \eta\tau_\pi\omega^2 - \left({\kappa\over 2}-\kappa^*\right)\omega^2 
- {\kappa\over 2}k_z^2 
\label{eq:derivative_expansion}
\end{align}
where $\eta$ is the shear viscosity, $\tau_\pi$ is the corresponding relaxation time,
and $\kappa$ and $\kappa^*$ are additional transport coefficients, and we have omitted
the higher order terms.\footnote{%
We follow the convention where only the non-zero components of the arguments are
displayed in a function of a 4-vector.
Thus, $G_R^{xy,xy}(\omega, k_z)$ means that $k^\mu = (\omega,0,0,k_z)$.
}
At first glance, this looks like a reasonable representation of the shear response
function in the small $\omega$ and small $k$ limit.
However, if the two limits do not commute,
one must specify whether this expansion assumes $k > \omega$ or $k < \omega$.

Careful inspection of the derivation of Eq.(\ref{eq:derivative_expansion})
reveals that the expansion above is perfectly valid,
which seems to indicate that the two limits do commute, at least for this specific case.
Hence, the following alternate form of the Kubo formula should work
\begin{align}
\eta & = 
i\lim_{k_z\to 0}\lim_{\omega\to 0}\partial_\omega G_R^{xy,xy}(\omega, k_z)
\label{eq:alt_kubo_G2}
\end{align}
as well as the usual one with the $k_z\to 0$ limit taken first.
Eq.(\ref{eq:alt_kubo_G2}) cannot be valid if $G^{xy,xy}_R(\omega, k_z)$ possesses
a diffusive hydrodynamic pole.
Therefore, validating Eq.(\ref{eq:alt_kubo_G2}) is 
equivalent to showing that the correlation function $G^{xy,xy}_R(\omega, k_z)$ 
does not possess any hydrodynamic pole. This is indeed the case 
\cite{Denicol:2011fa,Romatschke:2015gic,Hong:2010at}
ultimately because variation only in the $z$ direction cannot induce diffusion in the
transverse ($x,y$) directions.

This may be taken as a clue that if the full analytic structure of the retarded
correlation functions of stress-energy tensors are revealed, there may be other similar
formulae as yet unknown in the limit where $\omega\to 0$ is taken first.
This is what we would like to explore in this work.
Along the way, we also would like to address another puzzle 
that appears in the usual
calculations of the viscosities using the skeleton expansion: it does not seem to matter
which limit is taken first 
even when Eq.(\ref{eq:derivative_expansion}) is not directly used.

In recent gravity-hydrodynamics 
literature, there appeared a few Kubo-like formulae where
some transport coefficients are obtained in the limit where $\omega\to 0$ is taken first
\cite{York:2008rr,Moore:2010bu,Moore:2012tc,Kleinert:2016nav}
(In this work, we will limit the analysis only to the 2-point functions.).
Most of these formulae concern thermodynamic transport coefficients
that have more to do with the static properties of the system than the dissipative
properties.
This is understandable given that
the zero-frequency limit of
the retarded correlation function coincides with the static Euclidean correlation function.
We will show below that these are faithfully reproduced within our analysis.

\section{Analytic structure of stress-energy tensor correlation functions}

The purpose of this section is to explain briefly 
how to obtain the most general analytic structures
of the stress-energy tensor correlation functions
in the small $\omega$ and small $k$ limit.
The main tool we use is the energy-momentum conservation laws, supplemented
by the consistency with the gravity-hydrodynamics results.
Including additional conserved currents is left for future work.

\subsection{Ward Identity Analysis}

In an isotropic and static medium, the following Ward identities can be derived by
taking two functional derivatives of the curved-space partition function with respect to
the Riemannian metric, and then analytically continuing to the Minkowski space
\cite{Deser:1967zzf,
Czajka:2017bod,
Herzog:2009xv,
Policastro:2002tn}
\begin{align}
0 & =
k_\alpha
\left(
\tilde{G}^{\alpha\beta,\mu\nu}(\omega, {\bf k})
{-} \eta^{\beta\mu}\langle\hat{T}^{\alpha\nu}\rangle
{-} \eta^{\beta\nu}\langle\hat{T}^{\alpha\mu}\rangle
{+} \eta^{\alpha\beta}\langle\hat{T}^{\mu\nu}\rangle
\right)
\nonumber \\
\label{eq:WardIds0}
\end{align}
where $\hat{T}^{\mu\nu}$ is the stress-energy tensor.
Because $\tilde{G}^{\alpha\beta,\mu\nu}(\omega,{\bf k})$
can contain possible contact terms,
it is not exactly the same
as the retarded correlation function, which is defined as
\begin{align}
G^{\alpha\beta,\mu\nu}_R(\omega, {\bf k})
=
-i\int d^4 x\, e^{i\omega t - i{\bf k}\cdot{\bf x}}
\theta(t) \langle{[\hat{T}^{\alpha\beta}(t,{\bf x}), \hat{T}^{\mu\nu}(0)]}\rangle
\label{eq:G_R_abcd}
\nonumber \\
\end{align}
The flat-space metric we use is $\eta_{\mu\nu} = \hbox{diag}(-1,1,1,1)$. 
Here, $\langle{\hat{\cal{O}}}\rangle = (1/Z)\hbox{Tr}\,(e^{-\beta\hat{H}}\hat{\cal{O}})$ is the thermal
average of the operator $\hat{\cal{O}}$.
Using the CPT properties of the stress-energy tensor operators, one can show that
$G_R^{\alpha\beta,\mu\nu}(\omega,{\bf k}) = G_R^{\mu\nu,\alpha\beta}(\omega,{\bf k})$.

The equivalent Ward identities
involving the retarded correlation functions
can be obtained by requiring that
the total energy-momentum is conserved, and 
the correlation functions are well-behaved in the $k \to 0$ limit.
The identities are 
\cite{Kovtun:2012rj,
Policastro:2002tn}
\begin{align}
\omega G_R^{00,0\mu} & = k_i \left(G_R^{0i,0\mu} 
+ \eta^{i\mu}(\varepsilon_0 + P_0)\right)
\nonumber \\
\omega G_R^{0j,0\mu} & = k_i \left( G_R^{ij,0\mu} 
- \eta^{0\mu} \eta^{ij}(\varepsilon_0+P_0)\right)
\nonumber \\
\omega G_R^{0j,lm}
& =
k_i \left(
\tilde{G}^{ij,lm}
- \left(\eta^{il}\eta^{jm} + \eta^{im}\eta^{jl} - \eta^{ij}\eta^{lm}\right)P_0 \right)
\nonumber \\
\omega G_R^{00,ij} & = k_l G_R^{0l,ij}
\label{eq:WardIds}
\end{align}
where $\varepsilon_0$ is the equilibrium energy density and $P_0$ is the equilibrium
pressure. From now on, we will use the enthalpy density
$h_0 = \varepsilon_0 + P_0$ for this combination.
The only exception here is 
$\tilde{G}^{{ij},{lm}}$
for which we have not yet
found the corresponding $G_R^{{ij},{lm}}$. Fortunately, it is not strictly necessary to do so
as this correlator does not directly involve the energy-momentum density.
From this expression, it is clear that the imaginary part of the retarded correlators
obeys the same identities but without the $\varepsilon_0$ and $P_0$
contact terms on the right hand side.

Using the above Ward identities, all correlators involving the energy-momentum
density operator $\hat{T}^{0\mu}$ can be expressed in terms of
$\tilde{G}^{{ij},{lm}}(\omega, {\bf k})$.
In a spatially isotropic system,
this correlator can be decomposed in terms of the five orthogonal rank-2 tensors involving
$\delta_{ij} = \eta^{ij}$ and $k_i k_j$
\cite{Czajka:2017bod} as follows
\footnote{
This expression first appeared in an unpublished note by L.G.~Yaffe.
}
\begin{align}
\lefteqn{
\tilde{G}^{{ij},{lm}}(\omega, {\bf k})
= } &
\nonumber \\ &
P_0(\delta_{il}\delta_{jm} + \delta_{im}\delta_{jl} - \delta_{ij}\delta_{lm})
\nonumber \\ &
+
G_1(\omega, k^2) \left(\hat{k}_i \hat{k}_l \hat{\delta}_{jm} + \hat{k}_i\hat{k}_m\hat{\delta}_{jl} 
+ \hat{k}_j \hat{k}_m \hat{\delta}_{il} + \hat{k}_j \hat{k}_l\hat{\delta}_{im}\right)
\nonumber \\ &
+
G_2(\omega, k^2) (\hat{\delta}_{il}\hat{\delta}_{jm} + \hat{\delta}_{im}\hat{\delta}_{jl} - \hat{\delta}_{ij}\hat{\delta}_{lm})
\nonumber \\ &
+
G_L(\omega, k^2) \hat{k}_i \hat{k}_j \hat{k}_l\hat{k}_m
\nonumber \\ &
+ 
G_{LT}(\omega, k^2)\left(\hat{k}_i \hat{k}_j \hat{\delta}_{lm} + \hat{\delta}_{ij}\hat{k}_l \hat{k}_m\right)
\nonumber \\ &
+
G_T(\omega, k^2)\hat{\delta}_{ij}\hat{\delta}_{lm}
\label{eq:Gss}
\end{align}
using $\hat{k}_i = k_i/k$ and $\hat\delta_{ij} = \delta_{ij} - \hat{k}_i\hat{k}_j$.
From now on, we will refer to the stress-energy tensor correlation functions as the
correlators and the functions appearing on the right hand side of Eq.(\ref{eq:Gss}) as the
response functions.
From Eq.(\ref{eq:Gss}), one can see that only $G_1, G_L$ and $G_{LT}$ can appear in 
Eq.(\ref{eq:WardIds}) because $G_2$ and $G_T$ are transverse to ${\bf k}$ in all indices.
See Appendix\,\ref{app:correlation_funcs} for details.

To further investigate the behaviours of the response functions,
consider first the small $\omega$ limit.
From Eqs.(\ref{eq:WardIds}) and (\ref{eq:Gss}), and using the fact that the left hand
side of Eq.(\ref{eq:WardIds}) must all vanish in the $k\to 0$ limit due to the
energy-momentum conservation, we get
\begin{align}
G_L(\omega,k^2)
& =
-h_0{\omega^2\over k^2} 
-
{\omega^4\over k^4}g_{L}(k^2) + O(\omega^5)
\label{eq:GL_small_omega}
\end{align}
where we defined
\begin{align}
g_{L}(k^2) = -G_R^{00,00}(0, {\bf k}) 
\end{align}
with $g_{L}(0) = Tc_v$. Here $c_v$ is the heat capacity, and 
$T$ is the temperature \cite{Czajka:2017bod}.
Similarly, for $G_{LT}$ and $G_1$,
\begin{align}
G_{LT}(\omega, k^2) 
& =
- h_0 {\omega^2\over k^2} - g_{LT}(k^2)\omega^2 +
O(\omega^3)
\label{eq:GLT_small_omega}
\end{align}
\begin{align}
G_1(\omega,k^2)
& =
-h_0{\omega^2\over k^2} - g_{1}(k^2)\omega^2 + O(\omega^3)
\label{eq:G1_small_omega}
\end{align}
where $g_{LT}(k^2)$ and $g_1(k^2)$ are yet to be examined.

\subsection{Analytic structures of $G_1$ and $G_L$}

We are now ready to write down the general form of the response functions $G_1$, $G_L$ and $G_{LT}$.
We must, of course, be guided by the physics of the many-body system.
Hence, we assume that in the small $\omega$ and $k$ limit, these response functions either contain
a diffusion pole or a sound pole. The possible existence of a branch cut
\cite{Moore:2018mma,Rocha:2024cge,Gavassino:2024rck} is not explored in the present study.
For $G_1$, consider a flow velocity given by ${\bf u} = (u^x(y),0,0)$. 
Then $\partial_x u_y \ne 0$ but all other derivatives vanish.

Since non-equilibrium perturbations in the local rest frame can be expressed as
$\sim \hat{T}^{ij}\partial_i u_j$
\cite{zubarev1974nonequilibrium, 
Hosoya:1983id}
the relevant response function is
$\tilde{G}^{{xy},{xy}}(\omega, k_y) = G_1(\omega, k_y^2)$.
Since this is a response to the disturbance in the orthogonal direction of the
hydrodynamic flow, $G_1$ represents the diffusive shear mode.

For the diffusive response function, the most general form that is compatible with
Eq.(\ref{eq:G1_small_omega}) is 
\cite{Kovtun:2012rj,
Czajka:2017bod,
Romatschke:2015gic,
Jeon:2015dfa,
Kovtun:2003vj}
\begin{align}
G_{1}(\omega, k^2)
& =
-
{\omega^2 (h_0 + k^2 g_{1}(k^2) + i\omega F_1(\omega, k^2))
\over k^2 - i\omega/D(\omega,k^2) - \omega^2 B(\omega, k^2)}
\label{eq:Gdiff}
\end{align}
All functions of $\omega$ and $k$ appearing in the analytic structures in this
work are of the form 
\begin{align}
F_1(\omega, k^2) = F_{1R}(\omega^2,k^2) - i\omega F_{1I}(\omega^2, k^2)
\end{align}
where $F_{1R}$ and $F_{1I}$ are real functions. This is because
the relevant spectral functions need to be odd in $\omega$ and even in $k$.
Three special cases are $D(\omega, k^2) = D_R(k^2) - i\omega D_I(k^2)$ 
appearing above, and
$Z(\omega, k^2)$ and $F_{T1}(\omega)$ appearing in $G_L, G_{LT}$, and $G_T$ below.
The numerator in Eq.(\ref{eq:Gdiff})
can accommodate any polynomial of $\omega$ and $k^2$
that is consistent with Eq.(\ref{eq:G1_small_omega}).
The properties of the functions appearing above will be explored in Section \ref{sec:fixing}.

The longitudinal response function $G_L$ can be obtained as
$G_L(\omega, k_z^2) = \tilde{G}^{zz,zz}(\omega, k_z)$.
This is relevant when the disturbance is in the same direction 
of the flow ($\partial_z u_z(z) \ne 0)$ and one is interested
in how that disturbance creates a pressure wave (sound) in the same direction.
The most general form of the sound pole structure compatible with
Eq.(\ref{eq:GL_small_omega}) is 
\cite{Kovtun:2012rj,
Czajka:2017bod,
Romatschke:2015gic,
Jeon:2015dfa}
\begin{align}
G_L(\omega, k^2)
& =
-
{\omega^2(h_0 + \omega^2 F_L(\omega, k^2))
\over
k^2 - \omega^2/Z(\omega,k^2) - i\omega^3 R(\omega, k^2)
}
\label{eq:Gsound}
\end{align}
where 
\begin{align}
Z(\omega, k^2) = Z_R(k^2) + \omega^2 Z_{R\omega}(k^2) - i\omega Z_I(k^2)
\label{eq:Z_omega_k}
\end{align}
Taking the small $\omega$ limit of this expression and comparing with
Eq.(\ref{eq:GL_small_omega}), we can see that
\begin{align}
g_{L}(k^2) = {h_0\over Z_R(k^2)} + k^2 F_{LR}(0,k^2)
\end{align}
which reveals that the functions $Z_R(k^2)$ and $F_{LR}(0,k^2)$ are thermodynamic.

The response function
$G_{LT}(\omega, k_z^2) = \tilde{G}^{xx,zz}(\omega, k_z)$
is relevant when the disturbance is in the same direction of the flow
and one is interested in how that disturbance creates a pressure wave (sound) in the orthogonal direction.
Hence, we can express it as
\begin{align}
G_{LT}(\omega, k^2)
& =
-
{\omega^2(h_0 + k^2 g_{LT}(k^2) + i\omega F_{LT}(\omega, k))
\over
k^2 - \omega^2/Z(\omega,k) - i\omega^3 R(\omega, k)
}
\label{eq:GLT_expression}
\end{align}
which is consistent with Eq.(\ref{eq:GLT_small_omega}).

To go further in figuring out the analytic forms of $G_2$ and $G_T$,
one needs some more physics guidance.
Physically, these functions have the following interpretations.
 
 In purely hydrodynamic studies,
 $G_2(\omega, k_z^2) = \tilde{G}^{xy,xy}(\omega, k_z)$
 does not appear as a response function since $\partial_x u_y(z) = \partial_y u_x(z) = 0$.
 It describes the evolution of the system in the $(x,y)$ direction
 when variation is only in the $z$ direction. As such, it cannot have a hydrodynamic pole.
 The response function $G_T(\omega, k^2) =
 (1/4)(\hat\delta_{ij}\hat\delta_{lm}\tilde{G}^{ij,lm}(\omega,{\bf k}))$
 is the correlation function of 
 $\hat{\delta}_{ij}\hat{T}^{ij} = 3\hat{P} - \hat{k}_i \hat{k}_j \hat{T}^{ij}$
 where $\hat{P} = (1/3)\hat{T}^{ii}$ is the pressure.
 Hence, $G_T$ is expected to have the sound pole.

 Since $G_2$ and $G_T$ are not accessible in purely hydrodynamic approach,
 we need to seek more information from gravity-hydrodynamics calculations where the
 perturbation is $\sim \hat{T}^{ij}h_{ij}$. Here $h_{ij}$ is the metric perturbation.

\subsection{Gravity-Hydrodynamics results}

In this section, we aim to clarify the pole structures of all five response functions
in Eq.(\ref{eq:Gss})
by using the solutions of the linearized equations of motion 
worked out in
\cite{Hong:2010at,
Romatschke:2015gic,
Moore:2010bu,
Moore:2012tc}.
We do not assume the conformal limit.

In this analysis, the metric perturbation is given by
\begin{align}
g_{\mu\nu}(x) \approx \eta_{\mu\nu} + h_{\mu\nu}(x),
\end{align}
The linearized energy-momentum conservation laws 
in the rest frame of a fluid cell are given by
\begin{align}
\partial_t \delta\varepsilon + h_0 \partial_i u^i = -{1\over 2}h_0 \partial_t h^i_i
\end{align}
and
\begin{align}
h_0 \partial_t u^i + \partial_j T^{ji} = -P_0 \partial_j h^{ji}
\end{align}
where the spatial stress tensor is given by
\begin{align}
T^{ji} & = (P_0 + \delta P + \Pi)\delta^{ij} + \pi^{ji} - P_0 h^{ji}
\end{align}
Indices are raised or lowered by the flat metric $\eta_{\mu\nu}$ and $\eta^{\mu\nu}$.
In particular, $h^{\mu\nu} = \eta^{\mu\alpha}\eta^{\nu\beta}h_{\alpha\beta}$
and $h^{\mu}_{\nu} = \eta^{\mu\alpha}h_{\alpha\nu}$.\footnote{%
One should note that the linear piece in
\begin{align}
g^{\mu\nu} = \eta^{\mu\nu} + \bar{h}^{\mu\nu}
\end{align}
is {\em not} $h^{\mu\nu}$.
Because $g^{\mu\alpha}g_{\alpha\nu} = \delta^{\mu}_{\nu}$, we actually have
\begin{align}
\bar{h}^{\mu\nu} = -h^{\mu\nu}
\end{align}
In this work, we will always use $h_{\mu\nu}$ and raise or lower the indices with the flat
space metric.
} 
Here, $u^i$ is the flow velocity component,
$\delta\varepsilon$ is the deviation of the
energy density from its equilibrium value, and $\delta P = v_s^2 \delta\varepsilon$ 
is the pressure deviation with the speed of sound defined by
$v_s^2 = \partial P/\partial\varepsilon$. 

The shear-tensor equation of motion is
\begin{align}
\pi^{ij} & = -\eta(\partial^i u^j + \partial^j u^i - (2\delta^{ij}/3)\partial_l u^l)
- \tau_\pi \partial_t \pi^{ij}
-\eta\partial_t h^{\langle ij\rangle}
\nonumber \\ &
+
\kappa\left(R^{\langle ij \rangle} - 2R^{0\langle ij \rangle 0}\right)
+ 2\kappa^* R^{0\langle ij \rangle 0}
\end{align}
where $\eta$ is the shear viscosity, $\tau_\pi$ is the shear relaxation time, and 
$\kappa$ and $\kappa^*$ are thermodynamic transport coefficients \cite{Baier:2007ix}.
The angular bracket around indices indicates the symmetric and traceless part.
The bulk-pressure evolution is governed by
\begin{align}
\Pi = -\zeta \partial_l u^l - \tau_\Pi\partial_t\Pi + \xi_5 R 
+ \xi_6 R^{00} - {\zeta\over 2}\partial_t h_{l}^{l}
\end{align}
where $\zeta$ is the bulk viscosity, $\tau_\Pi$ is the bulk relaxation time,
and $\xi_5$ and $\xi_6$ are additional thermodynamic
constants.
The linearized Riemann tensor is given by 
\begin{align}
R_{\alpha\mu\beta\nu}
=
{1\over 2}\left(
\partial_{\mu}\partial_{\beta} h_{\alpha\nu}
+
\partial_{\alpha}\partial_{\nu} h_{\mu\beta}
-
\partial_{\mu}\partial_{\nu} h_{\alpha\beta}
-
\partial_{\alpha}\partial_{\beta} h_{\mu\nu}
\right)
\label{eq:RiemannTensor}
\nonumber \\
\end{align}

In the linear response analysis, the energy-momentum tensor averages are given by
\cite{
Baier:2007ix,
Hong:2010at,
Romatschke:2015gic,
Moore:2010bu,
Moore:2012tc}
\begin{align}
\langle{\hat{T}^{\mu\nu}(\omega, k)}\rangle_h
& =
\left. {\partial \langle{T^{\mu\nu}}\rangle_h\over \partial h_{\alpha\beta}}\right|_{h_{\mu\nu} = 0}
h_{\alpha\beta}(\omega,k)
\nonumber \\ &
-
{1\over 2}G_R^{\mu\nu,\alpha\beta}(\omega, k)h_{\alpha\beta}(\omega, k)
+
O(h^2)
\end{align}
The first term is there because the equilibrium in a curved space is different 
from the equilibrium in the flat Minkowski space. This term is the same as the
counter terms in Eq.(\ref{eq:WardIds0}).

The goal here is to choose the metric perturbation and its spatial dependence judiciously
so that only a particular response function is singled out.
The details of the solutions will be given elsewhere.
In this section, we will just quote relevant results.
In all analyses below, the metric perturbation has the following form
\begin{align}
h_{ij}(t, x_l) = h_{ji}(t,x_l) = h_{ij}(\omega, k_l)e^{-i\omega t + i k_l x_l}
\end{align}
where $l$ is not summed.

Let's start with the one we already know something about.
From Eq.(\ref{eq:Gss}), we know that $G^{{xy},{xy}}(\omega, k_y) = G_1(\omega, k_y^2)$.
Therefore, to get $G_1$, we should use $h_{xy}(t,y) = h_{yx}(t,y)$
with all other metric components being zero.
Solving for
\begin{align}
\pi^{xy}(\omega, k_y) 
& =
-
\bar{G}_1(\omega, k_y^2)h_{xy}(\omega, k_y)
\end{align}
we can obtain
\begin{align}
\bar{G}_1(\omega, k^2)
& =
{- \omega^2( h_0 - i{\kappa - 2\kappa^*\over 2D_T}\omega )
\over
k^2 - i\omega/D_T - (\tau_\pi/D_T)\omega^2 } 
\end{align}
confirming that $G_1(\omega,k^2)$ has the diffusion structure. Here
$D_T = \eta/h_0$ is the shear diffusion constant.
The over-bar on the response function 
indicates that it is obtained by solving the
gravity-hydrodynamic equations up to the second order, and hence not as general as $G_1$
in Eq.(\ref{eq:Gdiff}).

From Eq.(\ref{eq:Gss}), we also know that 
$G_R^{{xy},{xy}}(\omega, k_z) = G_2(\omega, k_z^2)$.
Therefore, to get $G_2$, we should use $h_{xy}(t,z) = h_{yx}(t,z)$
with all other metric components set to zero.
Solving for 
\begin{align}
\pi^{xy}(\omega,k_z) =
-
\bar{G}_2(\omega, k_z^2)h_{xy}(\omega, k_z)
\end{align}
we can obtain
\begin{align}
\bar{G}_2(\omega, k^2)
& =
-
{i\omega\eta + {\kappa\over 2}(\omega^2 + k^2) - \kappa^* \omega^2 
\over 1 - i\omega\tau_\pi} 
\end{align}
This analytic form of $\bar{G}_2$
indeed justifies Eq.(\ref{eq:derivative_expansion}) as a well-defined expansion
since $\bar{G}_2(\omega, k^2)$ has no hydrodynamic pole. 

Comparing $\bar{G}_2$ with $\bar{G}_1$, we see that
\begin{align}
\bar{G}_2(\omega, k^2) = \bar{G}_1(\omega, 0)
-
{{\kappa\over 2}k^2\over 1 - i\omega\tau_\pi}
\end{align}
The most general form of $G_2$ (not $\bar{G}_2$) can then be written as
\begin{align}
G_2(\omega, k^2) = G_1(\omega, 0) - k^2 F_2(\omega,k)
\label{eq:G2_expression}
\end{align}
where $F_2$ is analytic around $\omega = 0$ and $k = 0$ but may contain poles on the
imaginary $\omega$ axis away from $\omega = 0$.

The purely longitudinal response function $G_L$ can be obtained
from Eq.(\ref{eq:Gss}), by choosing $\tilde{G}_R^{{zz},{zz}}(\omega, k_z)$.
In terms of the metric perturbation, this corresponds to
having $h_{zz}(t,z)$ as the sole non-zero component.
By solving for
\begin{align}
\delta T^{zz}(\omega, k_z) 
& =
-
{1\over 2}\bar{G}_L(\omega,k_z^2) h_{zz}(\omega, k_z)
\end{align}
we obtain
\begin{align}
\bar{G}_L(\omega, k^2)
= {N_L(\omega, k^2)\over S_{\hbox{\scriptsize sound}}(\omega, k^2)}
\label{eq:barGL}
\end{align}
The explicit expressions of the numerator $N_L$ and the denominator $S_{\hbox{\scriptsize sound}}$ 
can be found in Appendix \ref{app:details}. Here, we just note that
the numerator indeed has a sound pole and the whole expression reduces to that found 
in \cite{Hong:2010at}
in the conformal limit where $\zeta = 0$ and $\tau_\Pi = 0$.

The response function $G_{LT}$ is given by
\begin{align}
G^{{xx},{zz}}(\omega, k_z) = G_{LT}(\omega, k_z^2)
\end{align}
which can be also obtained using $h_{zz}(t,z)$ but solving for
\begin{align}
\delta T^{xx}(\omega, k_z)
& =
-
{1\over 2}\bar{G}_{LT}(\omega, k_z^2)h_{zz}(\omega, k_z)
\end{align}
The solution is
\begin{align}
\bar{G}_{LT}(\omega, k^2)
& =
{
N_{LT}(\omega, k^2)
\over 
S_{\hbox{\scriptsize sound}}(\omega, k^2)
}
\label{eq:barGLT}
\end{align}
The numerator $N_{LT}$ is given in Appendix \ref{app:details}.
Eq.(\ref{eq:barGLT}) is fully consistent with Eq.(\ref{eq:GLT_expression}).

The last one to consider is $G_T$.
From Eq.(\ref{eq:Gss}), one can get
\begin{align}
G^{{yy},{zz}}(\omega, k_x) + G^{{zz},{zz}}(\omega, k_x)
= 2G_T(\omega, k_x^2)
\end{align}
Hence, the perturbation to use is $h_{zz}(t,x)$
and we need to solve for
\begin{align}
\delta T^{yy}(\omega, k_x) + \delta T^{zz}(\omega, k_x)
& =
-
\bar{G}_T(\omega, k_x^2)h_{zz}(\omega, k_x)
\nonumber \\ &
\end{align}
The response function is of the sound-pole type
\begin{align}
\bar{G}_T(\omega, k^2)
=
{N_T(\omega, k^2)\over S_{\hbox{\scriptsize sound}}(\omega, k^2)}
\end{align}
The numerator $N_{T}$ can be found in Appendix \ref{app:details}.
The most general form of $G_T$ turns out to be
\begin{align}
G_T(\omega, k)
& =
-
{\omega^2(h_0 + i\omega F_{T1}(\omega))
\over
k^2 - \omega^2/Z(\omega,k) - i\omega^3 R(\omega, k)
}
\nonumber \\ &
-
{
k^2(k^2 g_{T2}(k^2) + i\omega F_{T2}(\omega, k))
\over
k^2 - \omega^2/Z(\omega,k) - i\omega^3 R(\omega, k)
}
\label{eq:GT_expression}
\end{align}
Note that the function $F_{T1}(\omega) = F_{T1R}(\omega^2) - i\omega F_{T1I}(\omega^2)$ does
not depend on $k$ because any possible $k$ dependence can be absorbed by $F_{T2}(\omega, k)$.
The response functions $\bar{G}_{LT}$ and $\bar{G}_T$ were not calculated in 
\cite{Hong:2010at}.

What we would like to do next is to relate the $\omega = k = 0$ values of functions
appearing in the expressions for the response functions
to physical transport coefficients.

\section{Constraints from the small $k$ conditions}
\label{sec:small_k_constraints}

By requiring that 
the coefficients of the various
$\hat{k}\hat{k}\delta$ terms in Eq.(\ref{eq:Gss}) are $O(k^2)$ 
and the coefficient of the $\hat{k}_i\hat{k}_j\hat{k}_l\hat{k}_m$ term is $O(k^4)$
in the small $k$ limit with $\omega \ne 0$, 
one can get three independent relationships between the response functions in
the small $k$ limit.
One of the $O(k^2)$ relationship is
\begin{align}
2G_1(\omega,k) - G_L(\omega, k) + G_{LT}(\omega, k) = O(k^2)
\end{align}
In the $k\to 0$ limit,
this can be thought of as an equation for $F_{LTR}(\omega^2, 0)$ and $F_{LTI}(\omega^2, 0)$
in $G_{LT}$ in terms of the functions appearing in $G_1$ and $G_L$. 
The equations are linear, but the solutions are fairly long and not very illuminating.
The $\omega\to 0$ limits are, however, relatively simple:
\begin{align}
F_{LTR}
& = 
{2h_0D_R\over Z_R}
\end{align}
and
\begin{align}
F_{LTI}
& =
-\frac{2 \left(D_R^2
   h_0 B_R+D_R F_{1R}-D_I h_0\right)}{Z_{R}}
   \nonumber \\ &
   +2 D_R h_0
   R_{R}+F_{LR}-\frac{2 D_R h_0 Z_{I}}{Z_{R}^2}
\end{align}
where we introduced a short-hand notation for the $\omega = k = 0$ values of the functions as
the name of the functions themselves. That is,
$D_R = D_R(0), B_R = B_R(0,0)$, etc., except for the response functions.

The second $O(k^2)$ relationship is
\begin{align}
2G_T(\omega,k) - G_{LT}(\omega, k) - G_L(\omega, k) = O(k^2)
\end{align}
Taking $k\to 0$, we can use this condition to put conditions on $F_{T1}(\omega)$ in $G_T$. 
In the zero $\omega$ and $k$ limit, we find
\begin{align}
F_{T1R} & = {1\over 2}F_{LTR}
\nonumber \\
F_{T1I} & = {1\over 2}( F_{LTI} + F_{LR} )
\end{align}

With $F_{LT}(\omega, 0)$ and $F_{T1}(\omega)$ thus determined, the remaining small $k$ condition,
\begin{align}
G_2 - 4G_1 + G_T + G_L - 2G_{LT} = O(k^4)
\end{align}
is fulfilled up to $O(k^2)$ with $\omega \ne 0$.
The $O(k^4)$ 
can then be achieved by solving for
$F_{T2R}(\omega^2,0)$ and $F_{T2I}(\omega^2,0)$ in $G_T$ to remove
the $O(k^2)$ terms.
The solutions are again long and not very illuminating, but the $\omega = k = 0$ pieces are simple
to write down:
\begin{align}
F_{T2R} 
& =
3h_0 D_R
\end{align}
and
\begin{align}
F_{T2I}
& =
\frac{F_{2R}-4 D_R^2 h_0}{Z_R} +2 g_{LT}
\nonumber \\ &
+3\left( h_0 (D_I - D_R^2 B_R) - D_R F_{1R} \right)
\end{align}

Similarly, if one considers the small $k$ behaviour of $\tilde{G}^{ij,lm}(0,{\bf k})$,
one can also deduce $G_2(0,k^2) = O(k^2)$, $G_T(0,k^2) = O(k^2)$ and
\begin{align}
G_2(0,k^2) + G_T(0,k^2) = O(k^4)
\end{align}
from which one can get
\begin{align}
g_{T2} + F_{2R} = 0
\end{align}

\section{Fixing the constants with second order hydrodynamics}
\label{sec:fixing}

We are now ready to relate the constants, namely the $\omega = 0$ and $k=0$ values of the
functions appearing in the response functions, to physical transport coefficients.
First, consider the denominators of the response functions that give the dispersion
relationships for diffusion and sound.
By comparing the denominator of $G_1(\omega, k^2)$ 
with the dispersion relationship of the momentum diffusion from the second order
hydrodynamics \cite{Czajka:2017bod},
one can show
\begin{align}
D_R & = D_T = {\eta\over h_0}
\nonumber \\
B_R D_R& = \tau_\pi
\label{eq:DRBR}
\end{align}
where $D_R$ can be recognized as the shear diffusion constant $D_T$.
By comparing the denominator of $G_L(\omega, k^2)$ and the sound dispersion relationship from second order
hydrodynamics, one can get
\begin{align}
Z_R & = v_s^2
\nonumber \\
Z_{R\omega} & = -\left( \tau_\pi\tau_\Pi v_s^2 + \tau_\Pi {4\eta\over 3h_0} +
\tau_\pi{\zeta\over h_0}\right)
\nonumber \\
Z_I & = \Gamma + v_s^2(\tau_\pi + \tau_\Pi)
\nonumber \\
R_R & = {\tau_\pi +\tau_\Pi\over v_s^2}
\nonumber \\
R_I & =
\frac{\tau_{\pi} \tau_{\Pi}-Z_{I} R_{R}}{v_s^2}
\label{eq:dispersion_sols}
\end{align}
Here $v_s$ is the speed of sound, and 
$\Gamma = (4\eta/3 + \zeta)/h_0$ is the sound attenuation constant.

Using the above, we can further determine
\begin{align}
F_{LTR} & = {2\eta\over v_s^2}
\nonumber \\
F_{LTI} 
& =
-\frac{2 \eta  \Gamma}{ v_s^4 }
+\frac{2(K - \eta \tau_\pi)}{v_s^2}+{F_{LR}}
\nonumber \\
F_{T1R} & = {\eta\over v_s^2}
\nonumber \\
F_{T1I} & = 
-\frac{\eta  \Gamma}{ v_s^4 }
+\frac{(K -\eta \tau_\pi)}{v_s^2}+{F_{LR}}
\nonumber \\
F_{T2R} & = 3\eta
\nonumber \\
F_{T2I} & =
\frac{F_{2R}-\frac{4 \eta ^2}{h_0}}{v_s^2}
+3(K - \eta\tau_\pi) +2 g_{LT}
\label{eq:constants1}
\end{align}
where 
$K = D_I h_0 - D_R F_{1R}$, $F_{LR}, F_{2R}$ and $g_{LT}$ are still unknown.

To go further, we need results from the gravity-hydrodynamics studies.
To compare with Eq.(\ref{eq:derivative_expansion}),
one can expand $G_2$ to get
\begin{align}
G_2(\omega, k^2)
& =
-i D_R h_0\omega 
   + B_R D_R^2 h_0\omega^2
   - K\omega^2
   \nonumber \\ &
   - F_{2R} k^2 + O(\omega k^2)
\label{eq:G2_k_then_w_0}
\end{align}
The first two terms match the corresponding
terms in Eq.(\ref{eq:derivative_expansion}) upon using
Eq.(\ref{eq:DRBR}).
Matching the other two terms, we have
\begin{align}
F_{2R}
& =
{\kappa\over 2} = -g_{T2}
\\
K & =
{\kappa\over 2} - \kappa^*
\label{eq:DI_AR_rel}
\end{align}
For the sound modes, we can compare our $G_L$ with $\bar{G}_L$ in Eq.(\ref{eq:barGL}).
By requiring $G_L - \bar{G}_L$ to vanish up to and including $O(\omega^6)$ with $k\sim \omega$,
we get
\begin{align}
v_s^2 F_{LR}
& =
-\frac{2 \kappa }{3}+\frac{4 \kappa^*}{3}
+2 \xi_5-\xi_6
+ {h_0 Z_R'\over v_s^4}
\end{align}
As argued before, $F_{LR}$ is indeed thermodynamic in nature.

Similarly, comparing the gravity-hydro result $\bar{G}_{LT}$ in Eq.(\ref{eq:barGLT})
with $G_{LT}$ in Eq.(\ref{eq:GLT_expression}) gives a further condition
\begin{align}
g_{LT} 
& =
-\kappa +2 \kappa^* 
\label{eq:gepss}
\end{align}
Within the context of the second order hydrodynamics,
it is consistent to set $Z_R' = \lim_{k\to 0}{\partial Z_R(k^2)/\partial k^2} = 0$
(but not within the context of the third order hydrodynamics as can be seen below). 
This fixes all constants on the right hand sides of Eq.(\ref{eq:constants1}) in terms of
transport coefficients.

One can compare terms that are of higher order in $\omega$ and $k$
to determine $F_{1I}, B_I$, and $F_{2I}$ and the derivatives of other functions. But since
the results are not very illuminating and do not yield useful Kubo formulae,
we will not do so here.

\section{Third order conformal hydrodynamics and the relaxation times}

In the previous section, we analyzed response functions within the context of the
second order hydrodynamics. Recently, third order hydrodynamic results have been obtained
by several groups
\cite{deBrito:2023tgb,
Brito:2021iqr,
Jaiswal:2013vta,
Panday:2024hqp,
El:2009vj,
Ye:2024phs,
Younus:2019rrt,
Chattopadhyay:2014lya,
Diles:2019uft,
Diles:2023tau,
Grozdanov:2015kqa}.
In this section, we repeat some of the analysis so far to compare with the third order
results. For simplicity, we will restrict ourselves to the massless case which may be
referred to as the classical conformal theory.
Within the kinetic theory, $\Pi = 0$ and $\tau_\Pi = 0$, and the relevant equations from 
\cite{Ye:2024phs}
in the fluid cell rest frame are 
\begin{align}
\partial_t \pi^{ij}
& =
-{\pi^{ij}\over \tau_\pi} - \varphi_{-1|0}\sigma^{ij}
- \partial_l \xi^{lij}
\\
\partial_t \xi^{lij} 
& =
-{1\over \tau_\xi}\xi^{lij} 
-{3\over 7}\partial^{\langle l}\pi^{ij\rangle}
- \partial_k \zeta^{klij}
\\
\partial_t \zeta^{ijkl} 
& =
-{1\over \tau_\zeta}\zeta^{ijkl} 
-{4\over 9}\partial^{\langle i}\xi^{jkl\rangle}
\end{align}
where $\varphi_{-1|0} = 8T^4/5\pi^2$, and $\xi^{lij}$ and $\zeta^{ijkl}$ are the new spin-3
and spin-4 degrees of freedom, respectively.
These are still conformal because 
$\partial_l\xi^{lij} \sim \partial_l \partial^{\langle l}\pi^{ij\rangle}$ can be
decomposed as a linear combination of 
$\nabla^2 \pi^{\ij}$ and $\partial^{\langle i} \partial_l \pi^{lj\rangle}$.
Both are Weyl invariant \cite{Grozdanov:2015kqa}.
One can also get the same set of linearized equations from 
\cite{Panday:2024hqp} 
(up to $\xi^{lij}$), and a slightly different
set from 
\cite{deBrito:2023tgb}
(up to $\zeta^{ijkl}$) that contains
slightly different coefficients. The differences come from three different methods employed
in these studies to get $\xi^{lij}$ and $\zeta^{ijkl}$.
As this difference does not change the qualitative conclusion of this section, we will
only consider the above equations.

By analyzing the resulting linearized equations of motion for the diffusive mode
and comparing with $G_1$, we get
\begin{align}
D_R & = D_T
\nonumber \\
B_R D_R & = \tau_\pi + \tau_\xi + \tau_\zeta
\nonumber \\
D_I & = 
\frac{8}{35} \tau_\pi \tau_\xi 
+\frac{5}{21}  \tau_\xi \tau_\zeta 
+D_R (\tau_\xi +\tau_\zeta)  
\nonumber \\
B_I D_R
& =
\tau_\pi \tau_\xi 
+\tau_\pi \tau_\zeta 
+\tau_\xi \tau_\zeta 
- D_I B_R
\end{align}
Compared to the second order hydrodynamics results in Eq.(\ref{eq:DRBR}), we see that
the meaning of $B_R$ has changed and $D_I$ and $B_I$ are now determined.

Previously, $B_R D_R$ was just $\tau_\pi$ in the conformal limit.
Now it is the sum of all relaxation times. 
If one sets $\zeta^{ijkl} = 0$,
one can show that $B_R D_R = \tau_\pi + \tau_\xi$.
Hence, depending on how many additional modes are added,
$B_R D_R$ can be $\tau_\pi$, $\tau_\pi + \tau_\xi$,
or $\tau_\pi + \tau_\xi + \tau_\zeta$.
Therefore, the meaning of the Kubo formula 
\begin{align}
{1\over 2}\lim_{\omega\to 0}
\lim_{k_z\to 0}
\partial_\omega^2G^{{xy},{xy}}(\omega, k_z) 
=  B_R D_R \eta - K
\end{align}
obtained from Eq.(\ref{eq:G2_k_then_w_0})
does not seem to have a definite meaning as 
the meaning of $B_R$ changes as the order of hydrodynamics changes.
One can certainly evaluate the left hand side
in a given theory as was done in 
\cite{Czajka:2017bod}. 
However, how it is related
to the shear relaxation time is not so certain anymore.
There does exist a possibility that the changes in $K$ may compensate
the changes in $B_R$. However, since $K$ is purely thermodynamic in the second order
hydrodynamics, it does not seem likely that it will develop a $\tau_\zeta$ or $\tau_\xi$
dependence in the third order hydrodynamics.  

Analysis of the longitudinal mode gives a similar conclusion.
For instance, 
\begin{align}
v_s^2 R_R =
\tau_\pi + \tau_\xi + \tau_\zeta
\end{align}
behaves the same way as $B_R$,
and
$Z_R'$, which vanished within the second order hydrodynamics, is now
\begin{align}
  Z_R'& = \frac{1}{945} \tau_\xi (81 \tau_\pi+80 \tau_\zeta)
\end{align}

The results so far clearly indicate that the Kubo formulae involving the relaxation times
are {\em not}
stable against adding more relaxation equations for higher order hydrodynamics.
Hence, it is not so clear whether meaningful Kubo formulae for the relaxation times can
actually exist, at least from the 2-point correlation functions.\footnote{%
Refs.~\cite{Moore:2010bu,
Moore:2012tc}
contain a Kubo formula for $\tau_\pi$ in terms of a three-point
correlation function of the stress-energy tensor components,
which we have not yet analyzed in this context.
}
For this reason, 
we will not consider Kubo formulae involving
any relaxation times in the next section.

\section{New Kubo Formulae}
\label{sec:kubo_formulae}

With most of the constants now known,
we can derive new Kubo formulae. 
The novel feature of many of these formulae is that 
the $\omega \to 0$ limit is taken first and then the $k\to 0$ limit is taken,
in contrast to the usual requirement 
for the thermodynamic limit.
Known examples of such Kubo formulae are
the chiral magnetic conductivity and chiral vortical conductivity
\cite{Kharzeev:2009pj,
Amado:2011zx,
Landsteiner:2012kd,
Satow:2014lva}.
It is tempting to call this the static limit, but at this
point, it is not clear whether these new formulae can be actually calculated purely within
the Matsubara formalism with $\omega_n = 0$ due to the presence of the $\omega$ derivatives.

First, consider 
\begin{align}
\lefteqn{
\Delta_{ij,lm}
\tilde{G}^{{ij},{lm}}(\omega, k)
} &&
\nonumber \\
& =
20P
+ 8G_1(\omega,k^2)
+ 8G_2(\omega,k^2)
\nonumber \\ &
+ {4\over 3}\left(G_L(\omega, k^2) + G_T(\omega, k^2) - 2G_{LT}(\omega, k^2)\right)
\label{eq:Gpipi}
\end{align}
which is what is usually used to calculate the shear viscosity $\eta$.
Here, 
\begin{align}
\Delta_{ij,lm}
=\left(\delta_{il}\delta_{jm} + \delta_{im}\delta_{jl} - {2\over 3}\delta_{ij}\delta_{lm}\right)
\end{align}
isolates the spin-2 part of a rank-2 tensor.
The usual Kubo formula for the shear viscosity is
\begin{align}
\eta = {i\over 20}
\lim_{\omega\to 0}\lim_{k\to 0}
\Delta_{ij,lm}
\partial_\omega\tilde{G}^{{ij},{lm}}(\omega, k)
\label{eq:eta_k_first}
\end{align}
All five response functions contribute to this Kubo formula.

By using our analytic forms and reversing the order of the limits, we have an alternative
Kubo formula for $\eta$
\begin{align}
\eta = {i\over 12}
\lim_{k\to 0} \lim_{\omega\to 0}
\Delta_{ij,lm}
\partial_\omega\tilde{G}^{{ij},{lm}}(\omega, k)
\label{eq:eta_omega_first}
\end{align}
Only $G_2$ and $G_T$ contribute to this Kubo formula since all others vanish in the
$\omega\to 0$ limit with $k\ne 0$.
Note that  
deriving Eq.(\ref{eq:eta_omega_first})
is possible only because we have the analytic structures of all five response
functions.
Note also 
that even though both Eq.(\ref{eq:eta_k_first}) and Eq.(\ref{eq:eta_omega_first})
can be used to calculate $\eta$, the two orders of the limits do not commute.

Next, consider
\begin{align}
G_R^{0x,0x}(\omega, k_y)
& =
{k_y^2\over \omega^2} G_1(\omega, k_y^2) 
\end{align}
Taking a derivative with respect to $\omega$, we obtain 
\begin{align}
\lefteqn{
\partial_\omega G_R^{0x,0x}(\omega, k_y)
= 
} &
\nonumber \\
&
-iF_{1R}(0,k_y^2) 
- i{g_{1}(k_y^2)\over D_R(k_y^2)}
- i{h_0\over k_y^2 D_R(k_y^2)}
+ {2h_0\over k_y^4 D_R^2(k_y^2)}\omega  
\nonumber \\ &
+ O(\omega/k_y^2) + O(\omega^2)
\label{eq:alt_eta_0}
\end{align}
in the small $\omega < k_y$ limit.
From Eq.(\ref{eq:alt_eta_0}), an interesting Kubo formula for $\eta$ can be obtained
\begin{align}
{\eta\over h_0} = 
-i\lim_{k_y\to 0}\lim_{\omega\to 0}
{h_0\over 
k_y^2\partial_\omega G^{0x,0x}_R(\omega, k_y)
}
\label{eq:alt_eta}
\end{align}
Taking one more $\omega$-derivative, one similarly gets
\begin{align}
{\eta\over h_0} = 
\left(
\lim_{k_y\to 0}\lim_{\omega\to 0}
{2h_0
\over
\displaystyle
k_y^4 \partial^2_\omega G_R^{0x,0x}(\omega, k_y)
}
\right)^{1/2}
\label{eq:alt_eta2}
\end{align}
One can get isotropic forms
by substituting
$k_y \to k$ and $G_R^{0x,0x} \to G_R^{0i,0i}/2$
on the right hand sides.

These formulae are interesting because the denominator can be associated with the transport
cross-section $\sigma_{\rm tr}$ through
$ \eta/h_0 \sim l_{\hbox{\scriptsize mfp}}
\sim 1/(n_{\hbox{\scriptsize sc}}\sigma_{\hbox{\scriptsize tr}})$ 
where $l_{\hbox{\scriptsize mfp}}$ is the mean free path and 
$n_{\hbox{\scriptsize sc}}$ denotes the density of the scatterers.
In other words, putting a lower bound on $\eta/s$ 
\cite{Kovtun:2004de,Cohen:2007qr} 
is roughly equivalent to putting an
upper bound on the in-medium transport cross-section.
Thus, these formulae may provide intriguing possibilities
to explore the lower bound of $\eta/s$
in the context of finite temperature field theory.
However, the exact relationship between the denominators of 
Eqs.(\ref{eq:alt_eta}, \ref{eq:alt_eta2}) and the transport cross-section 
is yet to be clarified.

We can derive a few 
more Kubo formulae for $\eta$ and $\zeta$ all with the $\omega\to 0$ limit
taken first.
Noting $\tilde{G}^{xx,xx}(\omega, k_z) = P + G_2 + G_T$, we get
\begin{align}
\lim_{k_z\to 0}\lim_{\omega\to 0}
%
%
\partial_\omega \tilde{G}^{xx,xx}(\omega, k_z)
& = 
-4i\eta
\end{align}
Reversing the order of the limits, we have
\begin{align}
\lim_{\omega\to 0}\lim_{k_z\to 0}
%
%
\partial_\omega \tilde{G}^{xx,xx}(\omega, k_z) 
& = 
-i\left( \zeta + {4\eta\over 3} \right)
\end{align}

For the energy density and the pressure correlators, we have
\begin{align}
\lim_{k\to 0}\lim_{\omega\to 0}\partial_\omega G^{00,00}_R(\omega,{\bf k})
& =
-
{i\over v_s^4}\left(\zeta + {4\eta\over 3}\right)
\label{eq:Gepseps_omega_first}
\end{align}
\begin{align}
\lim_{k\to 0}\lim_{\omega\to 0}\partial_\omega G_R^{00,{ii}}(\omega,{\bf k})
= 
-
4 {i\eta\over v_s^2}
\label{eq:Geps3P_omega_first}
\end{align}
\begin{align}
\lim_{k\to 0}\lim_{\omega\to 0}\partial_\omega \tilde{G}^{{ii},{jj}}(\omega,{\bf k})
= -12i\eta
\label{eq:G3P3P_omega_first}
\end{align}
We can combine them to get
\begin{align}
\lim_{k\to 0}\lim_{\omega\to 0}
\partial_\omega G^{\Delta P\Delta P}(\omega,{\bf k})
= -i\zeta
\label{eq:zeta_omega_first}
\end{align}
where $\Delta\hat{P} = \hat{P} - v_s^2\hat\epsilon$ with 
$\hat{P} = \hat{T}^{ii}/3$ being the pressure operator and $\hat\epsilon = \hat{T}^{00}$ 
the energy density operator.
If one reverses the order of the limits, 
$\partial_\omega G_R^{00,00} \to 0$,
$\partial_\omega G_R^{00,ii} \to 0$,
and
$\partial_\omega \tilde{G}^{ii,jj} \to -9i\zeta$,
Hence we also have
\begin{align}
\lim_{\omega\to 0}\lim_{k\to 0}
\partial_\omega G^{\Delta P\Delta P}(\omega,{\bf k})
= -i\zeta
\label{eq:zeta_k_first}
\end{align}
Thus, like $G_2$, it does not matter which order the limits are taken. 
This is because in this combination, the numerator behaves like
$\sim (\omega^2 - v_s^2 k^2)$ for small $\omega$ and $k$ which compensates the sound
pole in the denominator. 
It is also interesting that both $\eta$ and $\zeta$ can be obtained from 
the imaginary part ({\it i.e.}~the spectral
density) of the pressure auto-correlation
function by taking different orders of the limits.

The following Kubo formulae for thermodynamic transport coefficients
are equivalent to 
similar formulae in 
\cite{Hong:2010at,Moore:2010bu,Moore:2012tc},
and hence serve as a good consistency check:
\begin{align}
\lim_{k\to 0}
{v_s^2\over 2}\partial_k^2 G_R^{00,00}(0,{\bf k})
& =
-v_s^2 F_{LR} 
\nonumber \\
& =
+\frac{2 \kappa }{3}
-\frac{4 \kappa^*}{3}
-2 \xi_5
+\xi_6
-{h_0 Z_R'\over v_s^4}
\nonumber \\
\label{eq:Kubo_QR}
\end{align}
and 
\begin{align}
\lim_{k\to 0}\partial_k^2 G_R^{00,{ii}}(0,{\bf k})
& =
-4 g_{LT} = +4(\kappa - 2\kappa^*)
\label{eq:Kubo_gepss}
\end{align}
and
\begin{align}
\lim_{k\to 0}\partial_k^2 G_R^{{ii},{jj}}(0,{\bf k})
& =
8 g_{T2} = -4\kappa 
\label{eq:Kubo_gT2}
\end{align}

\section{On diagrammatic calculations of viscosities}

Standard diagrammatic discussion of shear viscosity
via the Kubo formula Eq.(\ref{eq:eta_k_first})
starts with the one-loop diagram with bare propagators.
Upon realizing that this diagram does not produce a sensible answer and also that the
viscosity should inversely depend on the transport cross-section, one then introduces resummed
propagators to see that $\eta_{\hbox{\scriptsize 1-loop}} \propto 1/\Gamma_{\hbox{\scriptsize th}}$
where $\Gamma_{\hbox{\scriptsize th}}$
is the thermal width of the
propagator (equivalently, the imaginary part of the self-energy).
This inverse dependence then necessitates resummation of the ladder diagrams via
a Schwinger-Dyson equation for the dressed vertex.

What gets often overlooked in this process is that as long as 
$\Gamma_{\hbox{\scriptsize th}} \ne 0$, it does not matter whether the limit $k \to 0$ is taken first or the limit
$\omega\to 0$ is taken first. As we have shown in this paper,
only $\partial_\omega G_R^{xy,xy}(\omega,k_z)$ 
and $\partial_\omega G^{\Delta P\Delta P}(\omega, {\bf k})$ are independent of the order of the
limits.
All other response functions give different answers
when different orders of the limits are taken. For instance, Eq.(\ref{eq:eta_omega_first})
and Eq.(\ref{eq:G3P3P_omega_first}) are different from the usual Kubo formula for $\eta$
and $\zeta$.

It is then worth asking what really is being calculated with the usual skeleton
diagram expansion technique using the resummed propgators.
For simplicity, let us consider the one-loop diagram for
\begin{align}
\eta = i\lim_{\omega\to 0}\lim_{k_z\to 0}\partial_\omega G_R^{xy,xy}(\omega,k_z)
\label{eq:eta_Gxyxy_kz}
\end{align}
in the real scalar $\lambda\phi^4$ theory. 
Since
\begin{align}
{\rm Im}G_R^{xy,xy}(\omega, k_z)
& =
{1\over 2}(1 - e^{-\beta\omega})\sigma_{\color{red} >}^{xy,xy}(\omega, k_z)
\end{align}
where $\sigma_{\color{red} >}^{xy,xy}(\omega, k_z)$ is the 
{\color{red} greater} Wightman function, 
we only need to examine
\begin{align}
\sigma^{xy,xy}_{{\color{red} >, }\hbox{\scriptsize 1-loop}}(\omega, k_z)
& =
2\int {d^4 p\over (2\pi)^4}\, p_x^2 p_y^2
n(p^0{\color{red} + \omega})(1+n(p^0))
\nonumber \\ &
\rho_\phi(p^0+\omega, {\bf p}+k_z{\bf e}_z) \rho_\phi(p^0, {\bf p})
\label{eq:sigma_xyxy_1_loop}
\end{align}
where 
\begin{align}
\rho_\phi(p)
=
-i
\left(
{1\over (p^0 - i\Gamma_{\hbox{\scriptsize th}}/2)^2 - E_p^2}
-
{1\over (p^0 + i\Gamma_{\hbox{\scriptsize th}}/2)^2 - E_p^2}
\right)
\nonumber \\
\end{align}
is the spectral density for the resummed propagator with $E_p=\sqrt{{\bf p}^2+m^2}$.
The 4 poles of $\rho_\phi(p^0+\omega,{\bf p}+k_z{\bf e}_z)$ are at
$p^0
\approx
\pm i\Gamma_{\hbox{\scriptsize th}}/2 -\omega \pm (E_p+ p_z k_z/E_p)$.
Since $\Gamma_{\hbox{\scriptsize th}}/2 \ne 0$, it does not matter
to the pinching pole contribution
which limit is taken first.
We can just set $\omega = k = 0$.
This is precisely what we expected from $G_2$.

Now let's consider Eq.(\ref{eq:eta_k_first}). We now have
\begin{align}
\lefteqn{
{1\over 20}
\Delta_{ij,lm}
\sigma^{ij,lm}_{\hbox{\scriptsize 1-loop}}(\omega, {\bf k})
} &&
\nonumber \\
& =
\int {d^4 p\over (2\pi)^4}\, 
n(p^0{{\color{red}+\omega}})(1+n(p^0))
\nonumber \\ &
{1\over 30}
(
4{\bf p}^4 + 8 {\bf p}^2({\bf p}\cdot{\bf k}) + ({\bf p}\cdot{\bf k})^2 + 3{\bf k}^2 {\bf p}^2
)
\nonumber \\ &
\rho_\phi(p^0+\omega, {\bf p}+{\bf k}) \rho_\phi(p^0, {\bf p})
\label{eq:sigma_ss_1_loop}
\end{align}
The pole positions are at
$p^0 \approx \pm i\Gamma_{\hbox{\scriptsize th}}/2 -\omega \pm (E_p + {\bf p}\cdot{\bf k}/E_p)$ and again there is no
obstacle in just setting $\omega = k = 0$.
But that violates what we found in Eqs.(\ref{eq:eta_k_first}) and
(\ref{eq:eta_omega_first}) that the two orders of the limits do not commute. Therefore,
it is not clear whether Eq.(\ref{eq:sigma_ss_1_loop}) is evaluating Eq.(\ref{eq:eta_k_first})
or Eq.(\ref{eq:eta_omega_first}).
In fact, when $\omega = k = 0$, the right-hand-side of
Eq.(\ref{eq:sigma_ss_1_loop}) should be interpreted as
just the angle averaged version of Eq.(\ref{eq:sigma_xyxy_1_loop}).
Hence all calculations that started with Eq.(\ref{eq:eta_k_first}) but could be able to
set $\omega = k = 0$ without any obstacles may have all been actually using $G_2$ via
Eq.(\ref{eq:eta_Gxyxy_kz}).

For the bulk viscosity calculation,
the function $\partial_\omega G^{PP}$ does care about the order of the limits 
while ladder diagrams do not. 
Fortunately, from Eqs.(\ref{eq:zeta_omega_first}) and (\ref{eq:zeta_k_first}),
we see that it is permissible to just set $\omega = k = 0$
as long as the auto-correlation function of $\Delta\hat{P} = \hat{P} - v_s^2\hat\epsilon$ is used.
This may be implying that the ladder diagram resummation is only valid for this particular 
correlation function.

A question then arises: How do we use Kubo formulae 
that do depend on the order of the limits?\footnote{Except the thermodynamic ones such as
Eqs.(\ref{eq:Kubo_QR}, \ref{eq:Kubo_gepss}, \ref{eq:Kubo_gT2}).}
The answer seems to be that 
they can be used to calculate viscosities
only if some non-perturbative approximation results in hydrodynamic poles for the
correlation functions.
For instance, Eq.(\ref{eq:alt_eta}) 
makes sense only if an approximation of
$G_R^{0x,0x}(\omega, k_y) = (k_y^2/\omega^2)G_1(\omega, k_y^2)$ contains a diffusion
pole. 
For instance, the usual skeleton diagram resummation does not accomplish this, but 
the Anderson localization schemes can yield a diffusion pole by assuming that the
interaction is point-like
\cite{PhysRevB.22.4666,chu1993anderson,PhysRevD.77.027701,Giannakis:2006dc}.
Thus, further development of non-perturbative approximation schemes is necessary
to fully utilize the new Kubo formulae.

\section{Conclusions}

The main purpose of this work is
to elucidate the analytic structure of the stress-energy tensor correlation
functions in the small $\omega$ and small $k$ limit.
We succeeded in doing so by combining the energy-momentum conservation laws and the
results from the gravity-hydrodynamics calculations and the recently developed
third order hydrodynamics.

From this endeavour, we have drawn the following conclusions.
First, the response function often used in the calculation of the shear viscosity,
$G^{xy,xy}_R(\omega, k_z)$, does not in fact possess the shear diffusion pole.
In retrospect, it is almost obvious that this particular correlator
cannot possess the shear diffusion pole
since variation only in the $z$ direction cannot induce diffusion in the
transverse ($x,y$) directions.
On the other hand, $G_R^{xy,xy}(\omega, k_y)$ is the
full hydrodynamics response function that possesses a proper shear diffusion pole.
The usual calculations via skeleton expansion in fact correspond to using
$G_R^{xy,xy}(\omega, k_z)$ even though the isotropic version of the Kubo formula is used.
Similarly, skeleton expansion for the bulk viscosity makes sense only if 
the auto-correlation function of 
$\Delta\hat{P} = \hat{P} - v_s^2\hat\epsilon$ is used.

Second, the Kubo formulae for the relaxation times are unreliable because 
the result of the second order hydrodynamic analysis and the third order one do not
match. One should perhaps go back to the kinetic theory analysis to establish
what the Kubo formulae for the relaxation times actually mean.

Third, we have shown that the viscosities can also be calculated with the $\omega\to 0$ limit
taken first.  For instance, the pressure-pressure spectral density can give the bulk
viscosity or the shear viscosity depending on whether the $k\to 0$ limit is taken first 
or the $\omega\to 0$ limit is taken first. We have also derived some interesting Kubo
formulae where the shear viscosity to enthalpy ratio is inversely proportional to some
power of the correlation functions. These formulae may offer an intriguing possibility
to further study the $\eta/s$ bound in terms of the bound on the transport cross-section. 

Even though the $\omega\to 0$ limit is taken first, the new Kubo formulae are not really
thermodynamical or static because usually an $\omega$-derivative is involved.
Equivalently, these formulae use the imaginary part of the retarded correlation functions. 
In this regard,
the Kubo formula in Eq.(\ref{eq:alt_eta2}) is special because it uses the {\em real}
part of a retarded correlation function.
Although the second-order frequency derivative makes it non-trivial to connect
it to a purely Euclidean formulation (thus to simpler lattice calculations),
it may provide an intriguing possibility.
We should note that to use the new Kubo formulae fully, one needs an approximation
scheme where the resummed correlation functions actually exhibit hydrodynamic poles.
In this sense, investigation of new techniques
to calculate the explicit pole structure of the correlation functions would be important.

Finally, we would like to note that
for the sake of brevity, we have omitted many of the details in the derivations in this work. 
We will report them in a separate publication.

\begin{acknowledgments}
S.J.~acknowledges the support of the Natural Sciences and Engineering Research Council of Canada (NSERC),
[SAPIN-2024-00026], and thanks the support of Institute for Basic Science in Daejeon,
Korea, where the initial part of
this work was carried out. S.J.~also thanks C.~Gale, G.~Moore, and S.Y.~Kim for insightful discussions.
A.C.~is supported in part by the National Science Centre (Poland) under the research Grant No. 2021/43/D/ST2/01154 (SONATA 17).
J.H.~is supported by the National Research Foundation of Korea (NRF) grant funded by the Korean government (MSIT) (No. RS-2024-00342514).
\end{acknowledgments}

\newpage
\appendix

\section{Details of $\bar{G}_L, \bar{G}_{LT}$ and $\bar{G}_T$}
\label{app:details}

The numerator for $\bar{G}_L$ is given by
\begin{align}
\lefteqn{N_L(\omega, k^2) = } &
\nonumber \\
&
-
h_0
\bigg(
v_s^2 \omega^2
-i \tau_\pi v_s^2 \omega^3
-i \tau_\Pi v_s^2 \omega^3
-\tau_\pi \tau_\Pi v_s^2 \omega^4
\nonumber \\ &
-\frac{4 i \eta  \omega^3}{3 h_0}
-\frac{i \omega^3 \zeta }{h_0}
-\frac{4 \eta  \tau_\Pi \omega^4}{3 h_0}
-\frac{\tau_\pi \omega^4 \zeta }{h_0}
-\frac{2 \kappa  \omega^4}{3 h_0}
\nonumber \\ &
+\frac{4 \kappa^* \omega^4}{3 h_0}
+\frac{2 \omega^4 \xi_5}{h_0}
-\frac{\omega^4 \xi_6}{h_0}
+\frac{2 i \kappa  \tau_\Pi \omega^5}{3 h_0}
\nonumber \\ &
-\frac{4 i \kappa^* \tau_\Pi \omega^5}{3 h_0}
-\frac{2 i \tau_\pi \omega^5 \xi_5}{h_0}
+\frac{i \tau_\pi \omega^5 \xi_6}{h_0}
\bigg)
\end{align}
which is in fact independent of $k$ at this order.
The denominator exhibits the typical sound pole structure 
\begin{align}
\lefteqn{
S_{\hbox{\scriptsize sound}}(\omega, k^2) = }
&
\nonumber \\
&
k^2 v_s^2 -\omega^2
+
i \tau_\pi \omega^3
+i \tau_\Pi \omega^3
+\tau_\pi \tau_\Pi \omega^4
\nonumber \\ &
-\frac{4 i \eta  k^2 \omega}{3 h_0}
-\frac{i k^2 \omega \zeta }{h_0}
-i k^2 \tau_\pi v_s^2 \omega
-i k^2 \tau_\Pi v_s^2 \omega
\nonumber \\ &
-\frac{4 \eta  k^2 \tau_\Pi \omega^2}{3 h_0}
-\frac{k^2 \tau_\pi \omega^2 \zeta }{h_0}
-k^2 \tau_\pi \tau_\Pi v_s^2 \omega^2
\end{align}
This has exactly the same structure as $G_L$ in Eq.(\ref{eq:Gsound}).
It also coincides with the sound dispersion relationship
in the second order hydrodynamics. 
The correlator
$\bar{G}_L(\omega, k_z)$ 
coincides with the correlator $G^{zzzz}_R(0,k_z)$
calculated in 
\cite{Hong:2010at}
in the conformal limit
where $\zeta = 0$ and $\tau_\Pi = 0$.
The sound pole denominator is common to all sound response functions $G_L, G_{LT}$ and
$G_T$.

The numerator for $\bar{G}_{LT}$ is given by
\begin{align}
\lefteqn{
N_{LT}(\omega, k^2)
= } &
\nonumber \\
&
h_0
\bigg(
v_s^2 w^2
-i \tau_\pi v_s^2 w^3
-i \tau_\Pi v_s^2 w^3
-\tau_\pi \tau_\Pi v_s^2 w^4
\nonumber \\ &
+\frac{2 i \eta  w^3}{3 h_0}
-\frac{i w^3 \zeta }{h_0}
+\frac{\kappa  w^4}{3 h_0}
-\frac{2 \kappa^* w^4}{3 h_0}
\nonumber \\ &
+\frac{2 \eta \tau_\Pi w^4}{3 h_0}
-\frac{\tau_\pi w^4 \zeta }{h_0}
+\frac{2 w^4 \xi_5}{h_0}
-\frac{w^4 \xi_6}{h_0}
\nonumber \\ &
-\frac{i \kappa  \tau_\Pi w^5}{3 h_0}
+\frac{2 i \kappa^* \tau_\Pi w^5}{3 h_0}
-\frac{2 i \tau_\pi w^5 \xi_5}{h_0}
+\frac{i \tau_\pi w^5 \xi_6}{h_0}
\nonumber \\ &
-\frac{k^2 \kappa  v_s^2 w^2}{h_0}
+\frac{2 k^2 \kappa^* v_s^2 w^2}{h_0}
+\frac{i k^2 \kappa  \tau_\Pi v_s^2 w^3}{h_0}
\nonumber \\ &
-\frac{2 i k^2 \kappa^* \tau_\Pi v_s^2 w^3}{h_0}
+\frac{4 i \eta  k^2 w^3 \xi_5}{h_0^2}
-\frac{2 i \eta  k^2 w^3 \xi_6}{h_0^2}
\nonumber \\ &
+\frac{i k^2 \kappa  w^3 \zeta }{h_0^2}
-\frac{2 i k^2 \kappa^* w^3 \zeta }{h_0^2}
\bigg)
\end{align}
which is consistent with $G_{LT}$ in Eq.(\ref{eq:GLT_expression}). 

The numerator for $\bar{G}_{T}$ is given by
\begin{align}
\lefteqn{
N_T(\omega, k^2) = 
}&
\nonumber \\
&
-
h_0\bigg(
v_s^2 \omega^2
-i \tau_\pi v_s^2 \omega^3
-i \tau_\Pi v_s^2 \omega^3
-\frac{i \eta  \omega^3}{3 h_0}
-\frac{i \omega^3 \zeta }{h_0}
\nonumber \\ &
-\tau_\pi \tau_\Pi v_s^2 \omega^4
-\frac{\eta  \tau_\Pi \omega^4}{3 h_0}
-\frac{\tau_\pi \omega^4 \zeta }{h_0}
-\frac{2 i \tau_\pi \omega^5 \xi_5}{h_0}
\nonumber \\ &
+\frac{i \tau_\pi \omega^5 \xi_6}{h_0}
+\frac{2 \omega^4 \xi_5}{h_0}
-\frac{\omega^4 \xi_6}{h_0}
-\frac{\kappa  \omega^4}{6 h_0}
+\frac{\kappa^* \omega^4}{3 h_0}
\nonumber \\ &
+\frac{i \kappa  \tau_\Pi \omega^5}{6 h_0}
-\frac{i \kappa^* \tau_\Pi \omega^5}{3 h_0}
+\frac{3 i \eta  k^2 v_s^2 \omega}{h_0}
+\frac{3 \eta k^2 \omega^2 \zeta }{h_0^2}
\nonumber \\ &
+\frac{3 \eta  k^2 \tau_\Pi v_s^2 \omega^2}{h_0}
-\frac{2 k^2 \omega^2 \xi_5}{h_0}
+\frac{\kappa  k^2 v_s^2 \omega^2}{2 h_0}
+\frac{\kappa  k^2 \omega^2}{6 h_0}
\nonumber \\ &
-\frac{\kappa^* k^2 v_s^2 \omega^2}{h_0}
-\frac{\kappa  k^4 v_s^2}{2 h_0}
-\frac{i \kappa  k^2 \tau_\Pi v_s^2 \omega^3}{2 h_0}
-\frac{i \kappa  k^2 \tau_\Pi \omega^3}{6 h_0}
\nonumber \\ &
+\frac{2 i k^2 \tau_\pi \omega^3 \xi_5}{h_0}
+\frac{4 i \eta  k^2 \omega^3 \xi_5}{h_0^2}
-\frac{2 i \eta  k^2 \omega^3 \xi_6}{h_0^2}
-\frac{i \kappa  k^2 \omega^3 \zeta }{2 h_0^2}
\nonumber \\ &
+\frac{i \kappa^* k^2 \omega^3 \zeta }{h_0^2}
+\frac{i \kappa^* k^2 \tau_\Pi v_s^2 \omega^3}{h_0}
\nonumber \\ &
+\frac{i \kappa  k^4 \tau_\Pi v_s^2 \omega}{2 h_0}
+\frac{i \kappa  k^4 \omega \zeta }{2 h_0^2}
-\frac{4 i \eta  k^4 \omega \xi_5}{h_0^2}
\bigg)
\end{align}
which is consistent with $G_{T}$ in Eq.(\ref{eq:GT_expression}). 

\section{Correlation functions in terms of the response functions}
\label{app:correlation_funcs}

From Eqs.(\ref{eq:WardIds}) and (\ref{eq:Gss}), we can deduce
\begin{align}
 G_R^{0j,lm}(\omega, k)
& =
{k\over \omega}
\big(G_1(\omega, k^2)(\hat{k}_l\hat{\delta}_{jm} + \hat{k}_m\hat{\delta}_{jl})
\nonumber \\ &
+ G_L(\omega, k^2) \hat{k}_j\hat{k}_l\hat{k}_m
\nonumber \\ &
+ G_{LT}(\omega, k^2)\hat{k}_j\hat{\delta}_{lm}
\big)
\end{align}

\begin{align}
G_R^{0i, 0j}(\omega, k)
& =
{ k^2 \over \omega^2}
\left(G_1(\omega, k^2)\hat{\delta}_{ij} 
+
G_L(\omega, k^2)\hat{k}_i\hat{k}_j
\right)
\end{align}

\begin{align}
G_R^{00,ij}
& =
{k^2\over\omega^2}
\left( G_L(\omega, k^2)\hat{k}_i\hat{k}_j
+
G_{LT}(\omega, k^2)\hat{\delta}_{ij}
\right)
\end{align}

\begin{align}
\omega^3 G_R^{00,0i}(\omega, k)
& =
k_i k^2 \left( G_L(\omega, k^2) +{\omega^2\over k^2} h_0 \right)
\end{align}

\begin{align}
\omega^4 G_R^{00,00}(\omega,k)
& =
k^4\left(G_L(\omega, k^2) + {\omega^2\over k^2}h_0\right)
\end{align}

\begin{align}
\tilde{G}_R^{ii,jj}(\omega, k)
& =
3P
+ G_L(\omega, k^2)
+ 4G_{LT}(\omega,k^2)
+ 4G_T(\omega,k^2)
\nonumber \\ &
\end{align}

\begin{align}
\tilde{G}_R^{xy,xy}(\omega, k_y) = -P + G_1(\omega, k_y^2) 
\end{align}

\begin{align}
\tilde{G}_R^{xy,xy}(\omega, k_z) = -P + G_2(\omega, k_z^2)
\end{align}

\begin{align}
\tilde{G}_R^{yy,zz}(\omega, k_x) + \tilde{G}_R^{zz,zz}(\omega, k_x) = 2G_T(\omega, k_x^2)
\end{align}

\begin{align}
G^{00,xx}(\omega, k_z)
= {k_z^2\over\omega^2}G_{LT}(\omega, k_z^2)
\end{align}

\begin{align}
G^{0x,0x}(\omega, k_y)
& =
{k_y^2\over \omega^2}G_1(\omega, k_y^2)
\end{align}

\end{document}